\documentclass[aps,prd,twocolumn,amsmath,showpacs,superscriptaddress,nofootinbib,nopreprintnumbers]{revtex4-1}

\usepackage{verbatim}
\usepackage[T1]{fontenc}
\usepackage[utf8]{inputenc}
\usepackage[american]{babel}
\usepackage{epsfig}
\usepackage{graphicx}
\usepackage{booktabs}
\usepackage{multirow}
\usepackage{dcolumn}
\usepackage{amsmath}
\usepackage{mathtools}
\usepackage{amsfonts}
\usepackage{amssymb}
\usepackage{epstopdf}
\usepackage{bm}
\usepackage{siunitx}
\usepackage{braket}
\usepackage{enumitem}
\usepackage{soul}
\usepackage[table]{xcolor}
\usepackage{color}
\usepackage{transparent}
\usepackage{pifont}
\usepackage{todonotes}

\definecolor{navyblue}{rgb}{0.0, 0.0, 0.5}
\definecolor{royalblue}{rgb}{0.25, 0.41, 0.88}
\definecolor{cadmiumgreen}{rgb}{0.0, 0.42, 0.24}
\definecolor{blue-violet}{rgb}{0.54, 0.17, 0.89}
\definecolor{darkviolet}{rgb}{0.58, 0.0, 0.83}
\definecolor{orange(colorwheel)}{rgb}{1.0, 0.5, 0.0}

\usepackage{hyperref}
\hypersetup{
    colorlinks=true, 
    linkcolor=royalblue, 
    citecolor=magenta}

\newcommand{\mnu}{\sum m_\nu}

\usepackage{booktabs}
\usepackage{multirow}
\usepackage{dcolumn}
\usepackage{colortbl}

\definecolor{magenta(process)}{rgb}{1.0, 0.0, 0.56}

\definecolor{darkspringgreen}{rgb}{0.09, 0.45, 0.27}

\definecolor{royalblue(web)}{rgb}{0.25, 0.41, 0.88}

\begin{document}


\title{On the most constraining cosmological neutrino mass bounds} 

\author{Eleonora Di Valentino}
\email{eleonora.di-valentino@durham.ac.uk}
\affiliation{Institute for Particle Physics Phenomenology, Department of Physics, Durham University, Durham DH1 3LE, UK}

\author{Stefano Gariazzo}
\email{gariazzo@to.infn.it}
\affiliation{Istituto Nazionale di Fisica Nucleare (INFN), Sezione di Torino, Via P. Giuria 1, I-10125 Turin, Italy}

\author{Olga Mena}
\email{omena@ific.uv.es}
\affiliation{Instituto de F\'{i}sica Corpuscular (IFIC), University of Valencia-CSIC, Parc Cient\'{i}fic UV, c/ Cate\-dr\'{a}tico Jos\'{e} Beltr\'{a}n 2, E-46980 Paterna, Spain}

\date{\today}

\preprint{}
\begin{abstract}
We present here up-to-date neutrino mass limits exploiting the most recent cosmological data sets. By making use of the Cosmic Microwave Background temperature fluctuation and polarization measurements, Supernovae Ia luminosity distances, Baryon Acoustic Oscillation observations and determinations of the growth rate parameter, we are able to set the most constraining bound to date, $\sum m_\nu<0.09$~eV at $95\%$~CL. This very tight limit is obtained without the assumption of any prior on the value of the Hubble constant and highly compromises the viability of the inverted mass ordering as the underlying neutrino mass pattern in nature. The results obtained here further strengthen the case for very large multitracer spectroscopic surveys as unique laboratories for cosmological relics, such as neutrinos: that would be the case of the  Dark Energy Spectroscopic Instrument (DESI) survey and of the Euclid mission.

\end{abstract}
\maketitle
\section{Introduction} \label{sec:intro} 

Relic neutrinos with sub $eV$ masses represent a good fraction (if not all) of the hot dark matter component in our current universe. These hot thermal relics leave clear signatures in the cosmological observables, see e.g.~\cite{Lesgourgues:2018ncw,Lesgourgues:2006nd,Lattanzi:2017ubx,deSalas:2018bym,Vagnozzi:2019utt}, which can be exploited in order to put constraints on neutrino properties.
The most robust and constraining recent upper bound at $95\%$~CL on the total neutrino mass is $\sum m_\nu< 0.11$~eV~\cite{Aghanim:2018eyx}, obtained within the minimal $\Lambda$CDM framework. This limit can however be tightened or loosened depending on the data sets exploited in the analyses and also on the different assumptions concerning the remaining cosmological parameters, as, for instance, the dark energy equation of state, the curvature of the Universe or the possible presence of extra hot thermal relics, as sterile neutrinos or QCD axions~\cite{DiValentino:2015sam,Palanque-Delabrouille:2019iyz,Lorenz:2021alz,Poulin:2018zxs,Ivanov:2019pdj,Giare:2020vzo,Yang:2017amu,Vagnozzi:2018jhn,Gariazzo:2018meg,Vagnozzi:2017ovm,Choudhury:2018byy,Choudhury:2018adz,Gerbino:2016sgw,Yang:2020uga,Yang:2020ope,Yang:2020tax,Vagnozzi:2018pwo,Lorenz:2017fgo,Capozzi:2017ipn,DiValentino:2021zxy,DAmico:2019fhj,Colas:2019ret}. 

On the other hand, neutrino oscillations measured at terrestrial experiments indicate that at least two massive neutrinos exist in nature. Experiments, indeed, measure two squared mass differences, the atmospheric $|\Delta m^2_{31}| \approx 2.55\cdot 10^{-3}$~eV$^2$ and the solar $\Delta m^2_{21} \approx 7.5\cdot 10^{-5}$~eV$^2$  splittings~\cite{deSalas:2020pgw,Esteban:2020cvm}. 
Since the sign of $|\Delta m^2_{31}|$ is unknown, two possible mass orderings are possible, the \emph{normal} and the \emph{inverted} orderings. In the normal ordering, $\sum m_\nu \gtrsim 0.06$~eV, while in the inverted ordering, $\sum m_\nu \gtrsim 0.10 $~eV.
The above mentioned current cosmological limits are approaching to the minimum sum of the neutrino masses allowed in the inverted hierarchical scenario.
Cosmology can therefore help in extracting the neutrino mass hierarchy~\cite{Gariazzo:2018pei,RoyChoudhury:2019hls,Hannestad:2016fog,Lattanzi:2020iik}, which is a crucial ingredient in future searches  of neutrinoless double beta decay~\cite{Agostini:2017jim,Giuliani:2019uno}.

In this work, we shall update the cosmological constraints on neutrino masses from a number of current available cosmological measurements within the minimal $\Lambda$CDM scheme.
We shall add to CMB observations from Planck and to Supernovae Ia luminosity distance data the precious information from galaxy clustering surveys, as it is precisely in large scale structure where the free streaming nature of neutrinos plays a major role.
We shall exploit the results on Baryon Acoustic Oscillations (BAO) and Redshift Space Distortions from the spectroscopic SDSS-IV eBOSS survey~\cite{Dawson:2015wdb,Alam:2020sor}.
In the context of  spectroscopic observations, the BAO signature can be exploited in two possible ways. Along the line of sight direction, BAO data provide a redshift dependent measurement of the Hubble parameter $H(z)$.
Instead, across the line of sight, BAO data  can be translated into a measurement at the redshift of interest of the angular diameter distance, which is an integrated quantity of the expansion rate of the universe $H(z)$.
In addition, anisotropic clustering in spectroscopic BAO measurements can also be exploited to extract Redshift Space Distortions (RSD)~\cite{Kaiser:1987qv}.
This effect, due to galaxy peculiar velocities, modifies the galaxy power spectrum  and allows for an extraction of the product of the growth rate of structure ($f$) times the clustering amplitude of the matter power spectrum ($\sigma_8$), the well-known $f\sigma_8$ observable.

The structure of the paper is as follows. We describe the data used in the neutrino mass analyses in Sec.~\ref{sec:data}. The cosmological constraints on the total neutrino mass are presented in Sec.~\ref{sec:results}. We finish in Sec.~\ref{sec:concl} with a discussion of the impact of our results and future prospects.

\section{Methods} \label{sec:data} 
We describe in what follows the data and the analysis methods used here.
The most recent publicly available cosmological observations considered in deriving the neutrino mass bounds are: 

\begin{itemize}

\item {\bf Planck}: The Cosmic Microwave Background (CMB) temperature and polarization power spectra from the final release of Planck 2018 {\it plikTTTEEE+lowl+lowE}~\cite{Aghanim:2018eyx,Aghanim:2019ame}.

\item {\bf lensing}: The CMB lensing reconstruction power spectrum data obtained with a CMB trispectrum analysis~\cite{Aghanim:2018oex}.  

\item {\bf Pantheon}: Pantheon sample~\cite{Scolnic:2017caz} of the Type Ia Supernovae, consisting of 1048 data points, are also considered in the analysis.

\item {\bf BAO}: We exploit here the Baryon Acoustic  and Redshift Space Distortions measurements from SDSS spectroscopic galaxy and quasar catalogs. More precisely, we make use of the BOSS~\cite{Dawson:2012va} DR12 Luminous Red Galaxies (LRGs) results both in their \emph{BAO only} and in their \emph{BAO+RSD} forms~\cite{Alam:2016hwk}. Notice however that when combining DR12 with the eBOSS DR16 data we have only exploited the first two DR12 redshift bins in the $0.2 < z < 0.6$ region, (split into the $0.2 < z < 0.5$ and $0.4 < z < 0.6$ regions), as the more recent eBOSS DR16 results also analyzed here combine with BOSS results for z > 0.6, even if clearly the DR16 LRG results supersede the original BOSS DR12 results for the 0.5 < z < 0.75 redshift bin. 
\end{itemize}

The eBOSS survey~\cite{Dawson:2015wdb} started in 2014 for an operational period of five years. Targets for this spectroscopic survey were LRGs, Emission  Line  Galaxies  (ELGs), and Quasars (QSOs), and the cosmological interpretation of these observations can be found in Ref.~\cite{Alam:2020sor}. In our analyses we do not consider the measurements extracted from the ELGs observations, focusing exclusively in the cosmological constraints arising from the LRGs and QSOs samples, see Refs.~\cite{Bautista:2020ahg,Gil-Marin:2020bct,Hou:2020rse,Neveux:2020voa} for the detailed analyses from both the correlation function and the power spectrum. 

Concerning the LRG catalog, it covers the redshift range $0.6< z <1$ and, as previously described, it includes the BOSS DR12 LRGs within the $z >0.6$ tail of redshift distribution. We employ the measurements extracted from the eBOSS LRGs sample in their \emph{BAO only} form (i.e.\ via  $D_{\rm M}(z)/r_{\rm d}$ and $D_{\rm H}(z)/r_{\rm d}$ data) and also in their \emph{BAO+RDS} form (i.e.\ via $D_{\rm M}(z)/r_{\rm d}$, $D_{\rm H}(z)/r_{\rm d}$ and $f \sigma_8$ data), as we do with the BOSS DR12 LRG galaxies.

Concerning the eBOSS  QSO  sample, it spams over the $0.8< z <2.2$ redshift interval. As in the LRGs case, we shall exploit these observations in their \emph{BAO only} form and also in their \emph{BAO+RDS} form, albeit in the \emph{BAO+RDS} case the QSOs measurements of $D_{\rm M}(z)/r_{\rm d}$, $D_{\rm H}(z)/r_{\rm d}$  and $f \sigma_8$ are exclusively obtained by means of the RSD analyses, excluding the combination with the \emph{BAO only} measurements, see \cite{Alam:2020sor}.

The basic analysis is performed considering a $\Lambda$CDM+$\mnu$ model,
where the mass of the neutrinos is carried by three mass eigenstates if $\mnu>0.06$~eV, according to normal neutrino mass ordering, or by one mass eigenstate if $\mnu<0.06$~eV.
We use in the analysis the flat uniform priors on the parameters reported in Table~\ref{tab:priors}, and employ the Monte-Carlo Markov Chain code \texttt{CosmoMC}~\cite{Lewis:2002ah} a publicly free cosmological package (available from \url{http://cosmologist.info/cosmomc/}). The package supports the Planck 2018 likelihood~\cite{Aghanim:2019ame} having a precise convergence diagnostic based on the Gelman and Rubin statistics~\cite{Gelman:1992zz}. Additionally, this package appliances an efficient sampling of the posterior distribution which uses the fast/slow parameter decorrelations~\cite{Lewis:2013hha}. 
The cosmological observables are computed by \texttt{CAMB} \cite{Lewis:1999bs}, the Boltzmann solver associated with \texttt{CosmoMC}.
For most of our analyses, we consider three massive neutrinos in normal ordering if $\mnu>0.06$~eV or one massive neutrino if $\mnu<0.06$~eV, with a prior on $\mnu$ in $[0,5]~eV$.

\begin{table}
\begin{center}
\begin{tabular}{c|c}
\hline \hline
~~~~~~~~~Parameter~~~~~~~~~                    & ~~~~~~~~~Prior~~~~~~~~~\\
\hline 
$\Omega_{b} h^2$             & $[0.005,0.1]$\\
$\Omega_{c} h^2$             & $[0.001,0.99]$\\
$\tau$                       & $[0.01,0.8]$\\
$n_s$                        & $[0.8,1.2]$\\
${\rm{ln}}(10^{10}A_s)$         & $[1.6,3.9]$\\
$100\theta_{MC}$             & $[0.5,10]$\\ 
$\sum m_{\nu}~[eV]$               & $[0,5]$\\
$N_{\rm eff}$                & $[0.05,10]$\\
\hline
\hline
\end{tabular}
\end{center}
\caption{
Flat priors in the cosmological parameters analyzed here.
\label{tab:priors}
}
\end{table}

\begin{table*}
    \centering

    \resizebox{\textwidth}{!}{
    \begin{tabular}{|c|c|c|c|c|c|||c|c|}
        \hline
        \hline
	Planck+lensing & $\Sigma m_\nu$ & $H_0$
	&$\Omega_m$ & $\sigma_8$ & $S_8$& &\\ 
	+Pantheon &[eV] & [km/s/Mpc]
	&& & &$\ln B_{0-NH}$ &$\ln B_{NH-IH}$ \\ 
		\hline
	+ DR12 \emph{BAO only} & $<0.116$ & $67.8\pm1.0$ &  $0.309^{+0.013}_{-0.012}$ &  $0.814^{+0.017}_{-0.019}$ & $0.826\pm0.022$ & $-1.3$& $-1.5$\\
	+ DR12 \emph{BAO+RSD}& $<0.118$ & $67.8\pm1.0$ &  $0.310^{+0.013}_{-0.012}$ &  $0.814^{+0.017}_{-0.019}$ & $0.827^{+0.021}_{-0.022}$ & $-1.3$ & $-1.7$ \\
	+ DR16 \emph{BAO only}  & $<0.158$ & $67.5^{+1.2}_{-1.3}$ &  $0.314^{+0.017}_{-0.016}$ &  $0.811^{+0.020}_{-0.023}$ & $0.830^{+0.023}_{-0.024}$& $-0.7$ & $-1.6$ \\
	+DR16 \emph{BAO+RSD}  & $<0.101$ & $67.9^{+1.0}_{-1.1}$ &  $0.308^{+0.014}_{-0.013}$ &  $0.817^{+0.016}_{-0.017}$ & $0.828\pm0.022$& $-1.7$ & $-1.9$ \\
	+DR12 \emph{BAO only} + DR16 \emph{BAO only}  & $<0.121$ & $67.78^{+0.90}_{-0.97}$ &  $0.310^{+0.013}_{-0.011}$ &  $0.813^{+0.017}_{-0.019}$ & $0.826\pm0.021$ & $-0.9$ & $-1.8$ \\
	+DR12 \emph{BAO only} + DR16 \emph{BAO+RSD} & $<0.0866$ & $68.09^{+0.85}_{-0.88}$ &  $0.306\pm0.011$ &  $0.817^{+0.015}_{-0.016}$ & $0.826^{+0.020}_{-0.021}$& $-1.9$ & $-2.0$ \\
	+DR12 \emph{BAO+RSD} + DR16 \emph{BAO only} & $<0.125$ & $67.71^{+0.89}_{-0.97}$ &  $0.311^{+0.012}_{-0.011}$ &  $0.813^{+0.017}_{-0.019}$ & $0.828\pm0.021$& $-1.1$ & $-1.4$ \\
	+DR12 \emph{BAO+RSD} + DR16 \emph{BAO+RSD} & $<0.0934$ & $68.00^{+0.87}_{-0.89}$ &  $0.307\pm0.011$ &  $0.817^{+0.015}_{-0.016}$ & $0.827\pm0.021$& $-1.9$ & $-1.8$ \\
		\hline
    \end{tabular}}
    \caption{Constraints at 95\% CL on the total neutrino mass, the Hubble constant, the matter energy density in the universe, the  clustering amplitude of the matter power spectrum $\sigma_8$ and $S_8 \equiv \sigma_8 \sqrt{\Omega_m/0.3}$. In the last two columns we report the Bayes factor $\ln B_{0-NH}$ for the parametrization with $\sum m_\nu>0$~eV with respect to $\sum m_\nu>0.06$~eV, and $\ln B_{NH-IH}$ for the parametrization with $\sum m_\nu>0.06$~eV with respect to $\sum m_\nu>0.1$~eV, where negative values indicate a preference for the former cases.
    \label{tab:results}
    }
\end{table*}

\section{Results}
\label{sec:results} 

\begin{figure*}
\centering
\includegraphics[width=1.0\textwidth]{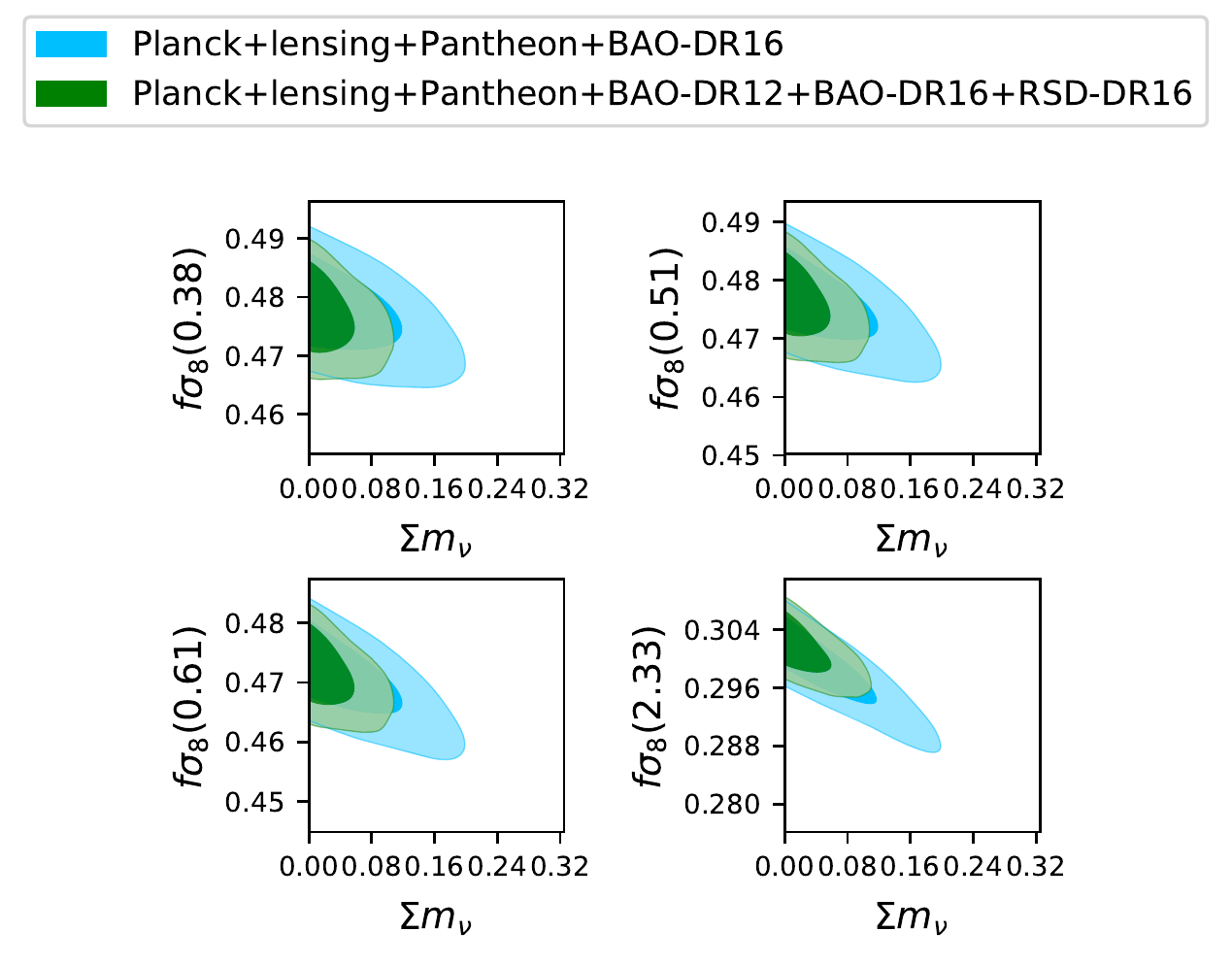}
\caption{
Values of the $f\sigma_8$ quantity, extracted via RSD measurements, at four different redshifts ($z=0.38,\,0.51,\,0.61,\,2.33$), versus the neutrino mass (in eV) for two different data sets explored here.
\label{fig:fsigma8}
}
\end{figure*}

Table \ref{tab:results} summarizes the main results of our analyses.
We consider a basic data combination, which includes CMB temperature, polarization and lensing, together with Pantheon Type Ia Supernovae luminosity and we add galaxy clustering information in a number of steps that we shall describe now. 
Firstly, we add BOSS DR12 measurements considering both \emph{BAO only} and \emph{BAO+RSD} information. In this case we also employ the information from the last redshift bin in the LRG DR12 sample, 
since we are not combining with DR16 measurements. The bounds we obtain are very similar, $\sum m_\nu <0.116$~eV and $\sum m_\nu <0.118$~eV, both at $95\%$~CL.
The situation is different when we consider eBOSS DR16 LRGs and QSOs samples. For these data combinations, the \emph{BAO+RSD} information is significantly more precious and constraining than the \emph{BAO only} one.
This can be understood in terms of Fig.~7 of Ref.~\cite{Alam:2020sor}: the value of $f\sigma_8$ for the QSOs sample at the mean redshift $z=1.48$ is much larger than the predictions obtained considering a $\Lambda$CDM universe with the best-fit parameters from Planck.
Since the addition of massive neutrinos further reduces the value of $f \sigma_8$, neutrino masses are severely constrained by DR16 QSOs \emph{BAO+RSD} measurements: we obtain $\sum m_\nu<0.1$~eV at $95\%$~CL when considering DR16 LRG plus QSOs \emph{BAO+RSD} information.
We illustrate this effect in Fig.~\ref{fig:fsigma8}, for two possible data sets at four different redshifts within the range of DR12 and DR16 measurements: notice that there is a clear reduction of $f\sigma_8$ as the neutrino mass is increased. 

\begin{figure*}
\centering
\includegraphics[width=1.0\textwidth]{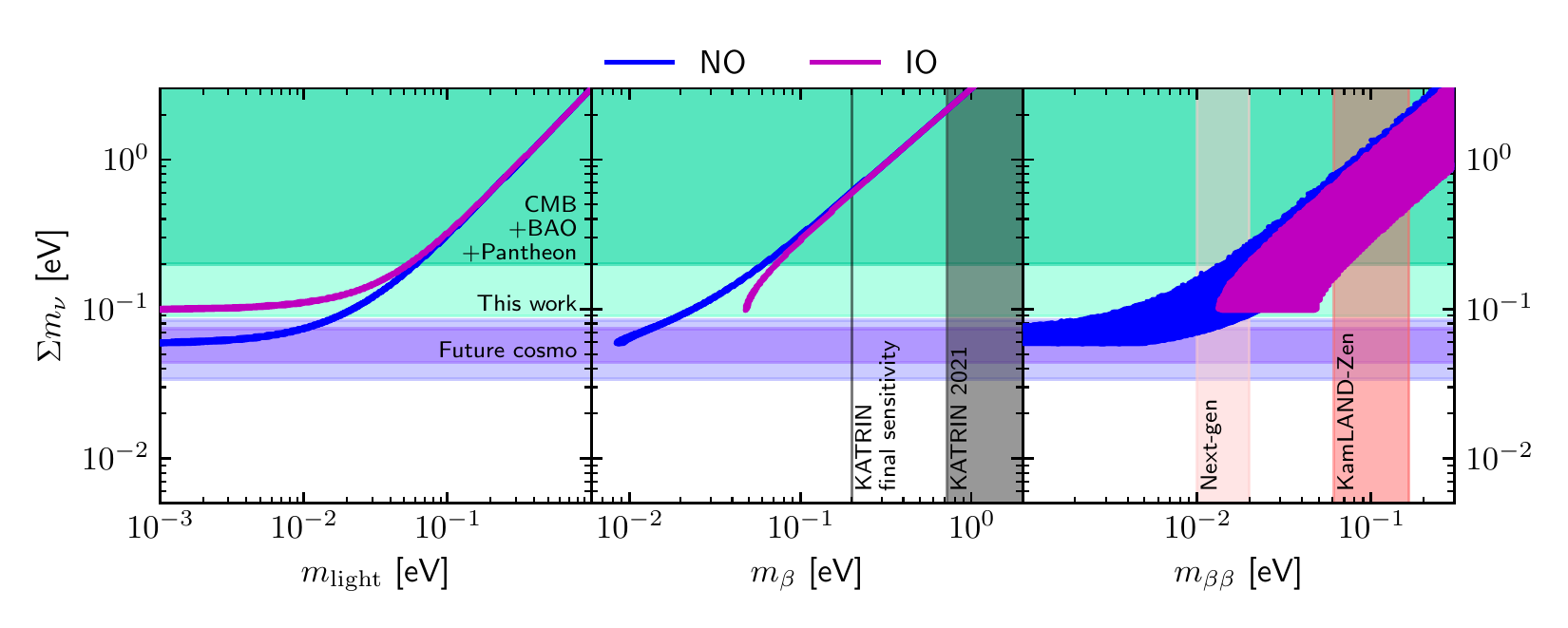}
\caption{ Sum of the neutrino masses ($\sum m_\nu$) as a function of other neutrino mass parameters:
lightest neutrino mass ($m_{\rm light}$, left panel), effective beta-decay mass ($m_\beta$, central panel) and effective Majorana mass relevant for neutrinoless double beta decay probes ($m_{\beta\beta}$, right panel).
Theoretical predictions for these quantities are depicted by the blue and magenta lines for normal (NO) and inverted ordering (IO), respectively.
Cosmological constraints from this paper are depicted by the green light band.
The future $1\sigma$ sensitivities from cosmological observations assuming a fiducial mass $\sum m_\nu=0.059$~eV are taken from Refs.~\cite{Basse:2013zua,Hamann:2012fe} (see purple regions).
The current and future $90\%$~CL sensitivities to $m_\beta$ can be found in Refs.~\cite{Aker:2021gma} and \cite{Drexlin:2013lha}, respectively.
Concerning neutrinoless double beta decay, current bounds constrain $m_{\beta \beta}<0.061-0.165$~eV at $90\%$~CL~\cite{KamLAND-Zen:2016pfg},
while next generation experiments are expected to reach $3\sigma$ sensitivities of $m_{\beta \beta}<0.010-0.020$~eV~\cite{Agostini:2017jim,Giuliani:2019uno}.
\label{fig:masslimits}
}
\end{figure*}

Furthermore, we combine BOSS DR12 and eBOSS DR16 measurements, in such case excluding the information from the last redshift bin in the LRG BOSS DR12 sample.
The \emph{less} constraining limit corresponds to the analysis of eBOSS DR16 data in their \emph{BAO only} form: $\sum m_\nu<0.12$~eV at $95\%$~CL.
The most constraining bounds are instead obtained when the eBOSS DR16 LRG and QSOs clustering information is taken in its \emph{BAO+RSD} form.
We obtain $\sum m_\nu<0.087$~eV and $\sum m_\nu<0.093$ at $95\%$~CL, depending on whether the BOSS DR12 LRG measurements are considered in their \emph{BAO+RSD} or \emph{BAO only} form, respectively.
To further assess the fact that the tightest constraints are driven by the QSOs RSD information, we have performed the following exercise: we have combined BOSS DR12 LRG data with the eBOSS DR16 LRG in its \emph{BAO+RSD} form, plus the QSOs in the \emph{BAO only} shape.
In this case we obtain a $95\%$~CL upper limit of $\sum m_\nu <0.12$~eV.
If we instead do the opposite, i.e.\ we analyse the eBOSS DR16 LRG sample in its \emph{BAO only} shape and the QSOs in the \emph{BAO+RSD} one, a very constraining bound is again recovered, i.e.\ $\sum m_\nu<0.09$~eV at $95\%$~CL.

Considering the number of relativistic degrees of freedom $N_{\rm{eff}}$ as a free parameter in the numerical analyses does not change dramatically the bounds.
For the combination of DR12 plus DR16 \emph{BAO+RSD} measurements, we obtain $\sum m_\nu<0.095$~eV and $N_{\rm{eff}}=3.05^{+0.33}_{-0.32}$, both at $95\%$~CL.
Moreover, if additional radiation is interpreted as sterile neutrino species, the constraints on the active neutrino masses are extremely stable.
Our limits in the three neutrino related parameters are $\sum m_\nu<0.09$~eV, $N_{\rm{eff}}<3.38$ and $m^{\rm{eff}}_{\nu, \rm{sterile}}<0.26$~eV (all at $95\%$~CL), arising from the analysis of DR12 plus DR16 \emph{BAO+RSD} data.
This is expected from the results of \cite{Gariazzo:2018meg}, where we can see that extending the cosmological model with a free $N_{\rm{eff}}$ does not alter significantly the constraints on $\sum m_\nu$.
References \cite{Gariazzo:2018meg, Gariazzo:2019xhx}
also show that one can expect the limits on $\sum m_\nu$ to 
be tightened \cite{Vagnozzi:2018jhn,Choudhury:2018byy}
or to
degrade by a significant amount (up to a factor $\sim2$)
when other cosmological parameters, such as the dark energy equation of state $w$ or the universe curvature $\Omega_k$, are varied as well.
A detailed analysis on the dependence of $\sum m_\nu$ on additional parameters, however,
is left for a future study.

It is also interesting to discuss the constraints in relation to the minimal allowed value of $\sum m_\nu$ by neutrino oscillations.
Cosmological measurements currently prefer values of $\sum m_\nu$ as close to zero as possible.
The peak of the posterior, therefore, is always below the neutrino oscillation values for such parameter.
The current bounds discussed above are already disfavoring the minimal allowed value for inverted ordering (IO) at more than 95\% CL,
but it is worth noticing that the minimal value for normal ordering (NO) is also (weakly) disfavored by cosmological measurements, as the strongest 68\% CL limit we obtain is $\sum m_\nu<0.037$~eV for the combination DR12 \emph{BAO} plus DR16 \emph{BAO+RSD} measurements.
A physically motivated prior on $\sum m_\nu$, therefore, cannot cover the region corresponding to higher likelihoods.
In order to quantify how much the laboratory prior on the total neutrino mass $\sum m_\nu>0.06$~eV is disfavoured by the cosmological data,
we compute the Bayesian evidence of a parametrization with $\sum m_\nu>0$ and one with $\sum m_\nu>0.06$~eV%
\footnote{
Notice that in this case we adopt the mass splittings obtained by neutrino oscillations in order to compute the sum of the neutrino masses as in~\cite{deSalas:2020pgw},
but as a baseline we adopt a Normal Ordering with $\mnu$ in the range $[0,5]$~eV, i.e.\ including the region excluded by laboratory experiments, where only one mass eigenstate is assumed. To perform the Bayesian comparison instead, we continue considering three neutrino masses in the Normal Ordering and restrict the prior for $\mnu$ to be above 0.06~eV, or we assume an Inverted Ordering and a prior above 0.1~eV.
Given the current precision of cosmological probes, the two methods are expected to be equivalent, see e.g.~\cite{Wang:2017htc},
while future probes may be able to distinguish the two cases, see e.g.~\cite{deSalas:2018bym}.
},
making use of the publicly available package \texttt{MCEvidence}%
\footnote{\href{https://github.com/yabebalFantaye/MCEvidence}{github.com/yabebalFantaye/MCEvidence}~\cite{Heavens:2017hkr,Heavens:2017afc}.}.
We compute the Bayes factor $\ln B_{ij}$ and report the results in the last two columns of Table~\ref{tab:results}.
We indicate with a negative (positive) value a preference of the cosmological data for parametrization with
$\sum m_\nu>0$~eV ($\sum m_\nu>0.06$~eV) in the $\ln B_{0-NH}$ case,
and for parametrization with $\sum m_\nu>0.06$~eV ($\sum m_\nu>0.1$~eV) in the $\ln B_{NH-IH}$ case.
We interpret the results with the revised Jeffreys scale~\cite{Kass:1995loi,Trotta:2008qt}, for which if $0 \leq | \ln B_{ij}|  < 1$ the evidence of the model is inconclusive, if $1 \leq | \ln B_{ij}|  < 2.5$ is weak, if $2.5 \leq | \ln B_{ij}|  < 5$ is moderate, and if $| \ln B_{ij} | \geq 5$ is strong.
When we perform a model comparison between the parametrization with $\sum m_\nu>0$ and the one with $\sum m_\nu>0.06$~eV, we obtain a weak evidence in favor of the first case if the RSD measurements are included in the dataset combination. In addition, we have a weak evidence for $\sum m_\nu>0.06$~eV (Normal Hierarchy) with respect to $\sum m_\nu>0.1$~eV (Inverted Hierarchy) for all the dataset combinations.
The strongest indication in favor of normal hierarchy is obtained when considering the DR12 BAO only plus DR16 BAO+RSD measurements,
with a Bayes factor $\ln B_{NH-IH}=-2$, corresponding to a 88\% probability (1.56$\sigma$) in favor of normal hierarchy being the true mass ordering.

Although this result is not yet statistically significant, if future data will robustly confirm the preference in favor of the $\sum m_\nu>0$ prior against the $\sum m_\nu>0.06$~eV one, we will have to face a serious tension between cosmology and neutrino oscillation experiments.

We also show some of the remaining cosmological parameters in Tab.~\ref{tab:results}.
Notice that the most constraining bounds on $\sum m_\nu$ lead to slightly smaller values of the present mass-energy density of matter $\Omega_m$: since the amount of the hot dark matter is reduced, the required cold dark matter component can also be reduced, as the suppression induced both in the CMB lensing and in the matter power spectrum grows as the value of $\sum m_\nu$ does.
Concerning the Hubble constant, the smaller the neutrino mass is, the larger should be $H_0$: the structure of CMB peaks fixes the value of $\Omega_m h^2$, and therefore, since a smaller neutrino mass implies a smaller value of $\Omega_m$, a mildly larger value of $H_0$ would be required.
The clustering parameters $\sigma_8$ and $S_8$ change by an insignificant amount.
Summarizing, the shift induced in the cosmological parameters shown in Tab.~\ref{tab:results} has no impact in the current cosmological tensions~\cite{DiValentino:2021izs,DiValentino:2020zio,DiValentino:2020vvd}.

\section{Conclusions} 
\label{sec:concl} 
We have analyzed current cosmological data from the Planck CMB  satellite, the SDSS-III and SDSS-IV galaxy clustering surveys and the Pantheon Supernova Ia sample to derive updated neutrino mass limits.
The most constraining limit we find here is $\sum m_\nu<0.09$~eV at $95\%$~CL, mostly due to Redshift Space Distortions analyses from the SDSS-IV eBOSS survey.
This upper limit lies below the minimum mass expected from oscillations in the inverted neutrino mass ordering, $\sum m_\nu\simeq0.10$~eV.
Such result has an enormous impact not only on theoretical models of neutrino masses, but also for ongoing and future kinematical searches for the neutrino mass.
Examples of these searches are represented by beta decay experiments, aiming at measuring the neutrino mass, or neutrinoless double beta decay probes, looking for the nature of neutrino masses, which can be Dirac or Majorana.
In Fig.~\ref{fig:masslimits} we illustrate, together with the theoretical expectations within each mass ordering, the current and future bounds for the three quantities characterizing the observables for neutrino masses: beta-decay ($m_\beta$), neutrinoless double beta decay $m_{\beta\beta}$ and the cosmological measured quantity $\sum m_\nu$.
The light green horizontal band depicts our most constraining bound, i.e.\ $\sum m_\nu<0.087$~eV at $95\%$~CL: this very tight limit has crucial implications for direct neutrino mass laboratory searches.
A model comparison between the parametrization with $\sum m_\nu>0$ and the one with $\sum m_\nu>0.06$~eV provides weak evidence in favor of the first case once the Redshift Space Distortions measurements are considered.
A weak preference in favor of the normal neutrino mass hierarchy versus the inverted one, instead, is almost independent of the considered data combination.
The very tight limits derived here arise from a combination of cosmological data that show no underlying tension and therefore should be regarded as extremely robust.
They make clear the huge relevance of galaxy clustering information for constraining particle properties and interactions.
Future spectroscopic measurements from ongoing and upcoming galaxy surveys are crucial for pinning down the neutrino mass limits from cosmological probes.
The Dark Energy Spectroscopic Instrument (DESI)~\cite{Levi:2013gra} has recently started its five years plan to survey the sky.
By providing the spectroscopic measurements of tens of millions of different tracers (Emission Line Galaxies, Luminous Red galaxies and Quasars) DESI can measure the sum of neutrino masses to sensitivities down to ~$0.02$ eV or better~\cite{Font-Ribera:2013rwa,Brinckmann:2018owf,Tanidis:2020byi}.
The future Euclid survey~\cite{Amendola:2012ys,Amendola:2016saw} will also have unique capabilities to constrain the neutrino mass and to break degeneracies with other unknown cosmological ingredients, such as the dark energy component~\cite{Basse:2013zua,Hamann:2012fe,Carbone:2011by}.

\begin{acknowledgments}
The authors would like to thank the precious help and support of H\'ector Gil-Mar\'in concerning eBOSS DR16 cosmological measurements.
EDV acknowledges the support of the Addison-Wheeler Fellowship awarded by the Institute of Advanced Study at Durham University.
SG acknowledges financial support from the European Union's Horizon 2020 research and innovation programme under the Marie Skłodowska-Curie grant agreement No 754496 (project FELLINI).
OM is supported by the Spanish grants FPA2017-85985-P, PROMETEO/2019/083 and by the European ITN project HIDDeN (H2020-MSCA-ITN-2019//860881-HIDDeN).
\end{acknowledgments}

\bibliography{biblio}
\end{document}